\documentstyle[natbib209,graphicx]{mn-nat}            %% MNRAS

\title[Photo-z Performance for Precision Cosmology]{Photo-z Performance for Precision Cosmology}
\author[Bordoloi et al.]{R. Bordoloi\thanks{E-mail:
rongmonb@phys.ethz.ch(RB)}, {S. J. Lilly} and {A. Amara}\\
Institute for Astronomy, ETH Z\"{u}rich, Wolfgang-Pauli-Strasse 27, CH-8093, Z\"{u}rich}
\begin{document}

\date{Accepted -Received -; in original form -}

\pagerange{\pageref{firstpage}--\pageref{lastpage}} \pubyear{-}

\maketitle

\label{firstpage}

\begin{abstract}

Current and future weak lensing surveys will rely on photometrically estimated redshifts of very large numbers of galaxies. In this paper, we address several different aspects of the demanding photo-z performance that will be required for future experiments, such as the proposed ESA Euclid mission.  It is first shown that the proposed all-sky near-infrared photometry from Euclid, in combination with anticipated ground-based photometry (e.g. PanStarrs-2 or DES) should yield the required precision in individual photo-z of $ \sigma_{z} (z) \leq 0.05(1+z)$ at $ I_{AB} \leq 24.5$.  Simple a priori rejection schemes based on the photometry alone can be tuned to recognise objects with wildly discrepant photo-z and to reduce the outlier fraction to $ \leq 0.25\%$ with only modest loss of otherwise usable objects.  Turning to the more challenging problem of determining the mean redshift $ \langle z\rangle$ of a set of galaxies to a precision of $ |\Delta_{\langle z \rangle}|\leq 0.002(1+z)$ we argue that, for many different reasons, this is best accomplished by relying on the photo-z themselves rather than on the direct measurement of $\langle z\rangle$ from spectroscopic redshifts of a representative subset of the galaxies, as has usually been envisaged. We present in an Appendix an analysis of the substantial difficulties in the latter approach that arise from the presence of large scale structure in spectroscopic survey fields.  A simple adaptive scheme based on the statistical properties of the photo-z likelihood functions is shown to meet this stringent systematic requirement.  We also examine the effect of an imprecise correction for Galactic extinction on the photo-z and the precision with which the Galactic extinction can be determined from the photometric data itself, for galaxies with or without spectroscopic redshifts. We also explore the effects of contamination by fainter over-lapping objects in photo-z determination.  The overall conclusion of this work is that the acquisition of photometrically estimated redshifts with the precision required for Euclid, or other similar experiments, will be challenging but possible.

\end{abstract}

\begin{keywords}
galaxies: distances and redshifts- cosmology: observations- methods: statistical.
\end{keywords}

\section{Introduction}

Large scale mapping of the weak lensing shear field in three dimensions is emerging as a potentially very powerful cosmological probe  \citep{Peacock2006, DETF}. Weak lensing has the advantage of directly tracing the mass distribution, thereby bypassing much of the complex astrophysics of the baryon component that underpin most of the other probes and which may well dominate the systematic uncertainties in them. In contrast, the underlying physics of weak lensing is extremely simple, and the challenges are primarily on the observational side, particularly the accurate measurement of the weak lensing distortion and the estimation of distances to very large numbers of faint galaxies.

A weak lensing survey of half the sky $ \rm (20,000\, deg^{2})$ to a depth of $ I_{AB} \sim 24.5$ and with a PSF of $ \sim$ 0.2 arcsec, forms a major part of the proposed ESA Euclid mission\footnote{http://www.ias.u-psud.fr/imEuclid \\ http://sci.esa.int/science-e/www/area/index.cfm?fareaid=102}. Euclid had its origins in two proposals submitted for the first round of the ESA Cosmic Visions 2015-2025 competition, the DUNE imaging survey \citep{DUNE2006} and the SPACE spectroscopic survey \citep{SPACE2009}.  The combination of the two surveys, plus the anticipated improved information on the Cosmic Microwave Background from Planck\footnote{www.rssd.esa.int/Planck} offers dramatic improvements in our knowledge of the entire dark sector, including the definition of the dark matter power spectrum, the dark energy equation of state parameter $ w$, as well as much else.

Application of weak lensing for cosmology requires at least a statistical knowledge of the distances, i.e. redshifts, of large numbers of individual galaxies. At $ I_{AB} <  24.5$, there are about 2.5 billion galaxies in the Euclid $ 2\pi$ sr survey area and so, realistically, reliance must be made on photometrically estimated redshifts (hereafter \textit{photo-z}).  

\subsection{Required redshift precision for precision cosmology with weak lensing} 

Several papers have discussed the redshift precision that is needed for weak lensing analyses to enable the full potential of this approach to be exploited \citep{Amara&Refrigier2007, Ma&HUetal2006, Abdalla2008}.

In the lensing tomography approach \citep{Hu_Tomography}, individual galaxies are binned into a number of redshift bins. The shear signal is extracted from the cross-correlation of the shape measurements of individual galaxies in different redshift bins.  These correlated alignments then give information (with some distance weighting function) on the mass distribution between the observer and the nearer of the two redshift bins.  Redshift information for the galaxies is required at two conceptually distinct steps: first, the construction of the redshift bins used for the cross-correlation analysis to extract the weak lensing signal and, second, the estimation of the mean redshift of the galaxies in a given bin, which is required to map the results onto cosmological distance and thereby extract the cosmological parameters. It is, of course, possible to do a similar correlation analysis with unbinned data \citep{2005PhRvD..72b3516C, kitching2008}, but for our purposes this distinction is unimportant. 

The cross-correlation between different redshift bins is undertaken to exclude any galaxy pairs that may be physically associated, i.e. have the same distance.  This is to avoid the possibility that physical processes operating around individual galaxies may produce an intrinsic alignment of the galaxies that may be mistaken for the coherent alignment produced by the weak lensing of the foreground mass distribution. The required accuracy of the individual photo-z for the bin-construction task is set by the need to exclude overlaps in the N(z) of individual bins (or the probability distribution for individual galaxies) and thereby remove physically close pairs \citep{2002A&A...396..411K}. This typically sets a requirement on the precision of individual photo-z of about $ \sigma_{z} = 0.05(1+z)$ \citep{bridle&King2007}.

There is a second type of intrinsic alignment effect \citep{2004PhRvD..70f3526H}, whereby the shape of the further of a given galaxy pair may be affected, through lensing effects, by the shape of the matter distribution around the nearer galaxy, which is likely to be correlated with the visible shape of that galaxy, thereby again producing a correlated alignment of the two galaxies that is unfortunately nothing to do with the lensing signal from the common foreground. \cite{Jochimi&Schneider2008},~\cite{Joachimi&Schnider2009} have shown that it is possible to implement a nulling approach to eliminate this second intrinsic alignment signal, which again requires a priori knowledge of individual redshifts.

Once the weak lensing signal is extracted, accurate knowledge of the redshift of the galaxies, as with any cosmological probe gives, amongst other parameters, information on the angular diameter distance $ D_{\theta}(z)$. The sensitivity of $ D_{\theta}(z)$ to the relevant cosmological parameters ($ \Omega_{m}$, $ \Omega_{\Lambda}$, $w$ etc) gives the required accuracy in the mean redshifts that are required to achieve a given precision on the parameters. As an example, \cite{Peacock2006} have shown that a precision of 1$ \%$ in $w$ requires a typical precision in the mean redshift of about 0.2$ \%$ in  $\langle z\rangle$. The Euclid goal is a 2$ \%$ precision in $w$ (independent of priors), requiring a precision of order 0.002(1+z) in $\langle z\rangle$. This simple approach is confirmed by extensive analysis of the Fisher matrices \citep{Amara&Refrigier2007, Ma&HUetal2006}. It is generally the case that if the mean redshift of a bin is defined accurately enough, then the higher moments of N(z) within the bin will also have been sufficiently determined. Of course, systematic biases in $\langle z\rangle$  that vary smoothly with redshift are particularly troublesome as they will mimic the effect of changing the cosmological parameters.

In summary, in order to reach the Euclid performance, we require a statistical (random) r.m.s. precision of order $0.05(1+z)$ per galaxy (for the correlation analysis), and a systematic precision in the mean z in each bin of order 0.002(1+z). These are both quite demanding requirements, and together with the shape measurement itself ($ \delta \gamma \sim 3 \times 10^{-4}$ - \citep{GREAT08, 2008MNRAS.391..228A}), they represent one of the observational challenges that lie along the path to enabling precision cosmology with weak lensing.  

Fortunately, there are some mitigating features of weak lensing analysis. For instance, the analysis is robust (aside from root-n statistics) to the exclusion of individual galaxies, provided only that the exclusion is unrelated to their shapes.  One is free therefore to reject galaxies that are likely to have poor photo-z provided that they can be recognized a priori, i.e. from the photometric data alone.

\subsection{Challenges for the spectroscopic calibration of N(z)}

Given the stringent requirements on the systematic error in the mean redshift $\langle z\rangle$ of a particular bin, one approach \citep{Abdalla2008} is to define the $N(z)$ and mean $\langle z\rangle$ through the acquisition of spectroscopic redshifts for a representative subset of the galaxies.  This direct sampling approach is certainly the most conservative, but will be very challenging, in practice, for the following reasons.

First, one clearly requires very large numbers of spectroscopic redshifts.  If we have a total redshift interval of $ \Delta z$, split into m bins, then the number of spectroscopic redshifts N will be of order:

\begin{equation}
N\, \sim  \, m^{-1} (\Delta z/ \sigma_{\langle z\rangle})^{2} 
\end{equation}
This assumes that the photo-z are perfect, and that there are no outliers.  This leads trivially \citep{Amara&Refrigier2007} to a requirement for $ 10^{5}-10^{6}$ spectroscopic redshifts.  

Secondly, these spectroscopic redshifts must be fully representative of the underlying sample.  Any biases in the sampling of the bin or, even harder to reliably quantify, the almost inevitable biases in the ability to secure a reliable spectroscopic redshift, must be dealt with via a possibly complex and inevitably somewhat uncertain weighting scheme \citep{limaetal2008}. It should be noted that current routine spectroscopic surveys of typical faint galaxies do not even approach 100$ \%$ completeness, even at brighter levels.  One of the best to date is the zCOSMOS survey \citep{Lilly2007} on relatively bright $ I_{AB} < 22.5$ galaxies which yields, at present, a 99$ \%$ secure redshift for 95$ \%$ of galaxies at its optimum $ 0.5 < z < 0.8$ \citep{Lilly2009}. Most other surveys are significantly less complete.

Even more invidious are the effects of large scale structure in the spectroscopic survey fields, often called cosmic variance. Our own semi-empirical analysis (see the Appendix) of the COSMOS mock catalogues \citep{kitzbichler&White2007} shows that, in a given patch of sky, the N(z) at $ I_{AB} \sim 24$ becomes dominated by cosmic variance as soon as a rather small number of galaxies have been observed.  The precise number depends on the field of view of the spectrograph, but is typically about 20-100 for spectroscopic fields of order 0.02-1 square degrees, i.e. a sampling rate of only a few percent.  This means that the spectroscopic survey must be split up over a very large number of independent fields and that to get $ 10^{6}$ redshifts that are Poisson variance dominated one must effectively cover the whole sky in a sparse sampled way.  This is unlikely to be efficient with the large telescopes needed for such faint object spectroscopy. A similar concern comes from the effects of Galactic extinction and reddening, which are likely, even when corrected for, to require spectroscopic sampling across the full range of Galactic [b,l].

These difficulties prompt consideration of other approaches, and in particular, that of placing greater reliance on the photo-z themselves, not only to construct the bins, but also to define their $\langle z\rangle$ with small systematic error.  

\subsection{Using photo-z for construction of N(z)} 

The performance of photometric redshifts is continually improving.  For example, in the COSMOS field \citep{2007ApJS..172....1S}, where we have very deep 30-band photometry from the ultraviolet(GALEX) to $ 5\mu m$, several photo-z schemes now achieve a precision of $\sigma_{z} \sim 0.01(1+z)$, with an outlier fraction (in non-masked areas and defined as a redshift difference greater than $0.15(1+z_{spec})$ of $ <1 \%$ \citep{2009ApJ...690.1236I}), at $ I_{AB} < 22.5$ and $ 0.05 < z < 1.4$, where the photo-z can be checked with about over 10,000 spectroscopic redshifts from zCOSMOS. It should be noted in passing that these 30 bands represent a very inhomogeneous data set in terms of point spread function, etc., and so this impressive photo-z performance also demonstrates the feasibility of combining disparate data into homogeneous photometric catalogues. Of course, this outstanding performance in the COSMOS field is unlikely to be achieved over the whole sky for the foreseeable future because of the expense of the required multi-band photometry. Nevertheless, the demonstration of this performance in COSMOS suggests that we have not yet reached any fundamental limit to photo-z performance.

There are a number of different approaches to photo-z estimation that can be broadly distinguished between template-matching and more purely empirical approaches, such as Artificial Neural Networks \citep{ANN_Lahav}. These have complementary strengths.  Template fitting is based on the observed limited dimensionality of galaxy spectral energy distributions plus an astrophysical knowledge of the effects that can modify them, e.g. the  the redshift itself, and the effects of extinction in our own Galaxy or in the distant galaxy.  The more empirical approaches in essence avoid any such assumptions, which is both a strength and a limitation. Although both approaches have passionate adherents, our own view is that both approaches can normally perform equally well and that both are normally limited by the quality of the available data.  In practice both use elements of the other, e.g. in template fitting, the actual data can be used to adjust the templates and the photometric zero-points, somewhat blurring the distinction.   Finally, it should be noted that both can produce a likelihood distribution in redshift space through the application of priors \citep{Hyperz, ANN_Lahav, BPZ, EASY}.  In this paper we will base our analysis on a template-fitting algorithm (ZEBRA,~\citep{2008arXiv0801.3275F}), since we believe its strengths are well suited to the problem in hand. We also note that the impressive real-life performance in COSMOS described above was achieved with two independent template fitting codes (Le Phare, \citep{2009ApJ...690.1236I} and ZEBRA \citep{Feldmann2006}). 

This improving photo-z performance described above suggests that it may be possible to use the photo-z themselves to construct the N(z), and thus $ \langle z\rangle$ for each bin, providing that the systematic uncertainties can be kept below the required level of 0.002(1+z). Some spectroscopic calibration would of course still be required, but the focus of this would be on constructing and characterizing the photo-z algorithm, rather than on constructing the N(z) directly.

This approach would have a number of potential advantages over that discussed by \cite{Abdalla2008} and others, and summarized above. At the very least, the number of spectra needed may be substantially reduced, although it is unlikely that one would wish to rely entirely on the photo-z and eliminate the spectroscopic conformation completely.  However, to characterize the uncertainties in the photo-z, defined by $ \sigma_{z}$, to the required level ($ \sigma_{\langle z\rangle}$) we would need of order 
\begin{equation}
N \sim  (\sigma_{z}/\sigma_{\langle z \rangle})^{2}
\end{equation}
spectroscopic redshifts, which may be orders of magnitude or more smaller than that implied by equation 1 since $ \sigma_{z} \sim 0.05 \Delta z$.

More importantly, the requirements on completeness and sampling are substantially relaxed since the photo-z characterization is done on individual objects.  As one example, it is relatively easy to simulate the degradation in photo-z performance with noisier photometric data, so the calibrating spectroscopic objects need not necessarily extend all the way down to the photometric limit.  The cosmic variance problem in spectroscopic calibration is eliminated, and the uncertainties arising from Galactic reddening can also be substantially reduced.

\subsection{Subject of this paper}

The aim of this paper is to explore the performance of a template fitting photo-z code as applied to simulated photometric data of the approximate quality that we may realistically expect for a $ 2\pi$ sr \textit{all-sky} ground and space survey within the next decade.  Our emphasis is on both the per object performance and on the potential for recognising and correcting systematic biases, which must be done to a high level if the increased reliance on photo-z is to be possible.

As described in more detail in Section 3, we will assume for definiteness a photometric data set that includes the three-band near infrared photometry that is planned for Euclid itself, plus 5-band grizy photometry similar to that which should be produced by the PanStarrs-1, -2 and -4 projects (hereafter PS-1, -2 and -4)\footnote{http://pan-starrs.ifa.hawaii.edu}. Examining these three generations of ground-based survey probes a range of depths that can be compared against other future surveys such as DES\footnote{http://www.darkenergysurvey.org} and LSST\footnote{http://www.lsst.org}. 

We then explore the following four topics that potentially may limit the photo-z performance and their usefulness to construct N(z) and $\langle z\rangle$:

\begin{enumerate}
\item The photo-z performance on individual objects in terms of the r.m.s. scatter (and bias) between the true redshift and the maximum likelihood photo-z, with particular emphasis on how to a priori identify and reject the outliers (\textit{catastrophic failures}) from their individual photo-z L(z) likelihood distributions. 
\item The construction of N(z) for a given set of photo-z selected galaxies, using their photo-z alone, with an emphasis on how to modify their individual likelihood functions L(z) to yield the least biased estimate of N(z) and $\langle z\rangle$ for the ensemble.
\item The systematic biases that can enter into the photo-z from an incorrect assumption about the level of foreground Galactic reddening, and how well the photometric data themselves can be used to determine the foreground reddening, both for a set of galaxies at known redshifts, and for those without known redshifts.
\item The effects of the photometric superposition of two galaxies at different redshifts, leading to a mixed spectral energy distribution that may perturb the photo-z, with an emphasis on seeing what happens to the redshift likelihood distribution. An interesting question is whether such composite objects can be recognised photometrically as well as ``morphologically'' from the images.
\end{enumerate}

Our approach is to try to isolate these problems and to explore each in turn with the aim of providing an \textit{existence proof} that provides a plausible route to achieve the very high photo-z performance that is required for Euclid. In particular, we decided to construct the input photometric catalogues using exactly the same set of approximately ten thousand templates as we subsequently used in the ZEBRA photo-z code.  This may strike some readers as being somewhat circular. However, this approach allows us to eliminate the choice of templates as a variable, or uncertainty, in our analysis.  This is motivated by the exceptional performance (discussed above) that has already been achieved with the same templates coupled with the exquisite observation data in COSMOS.  This strongly suggests to us that the choice of templates is unlikely to be the limiting factor with the degraded photometry that we can realistically expect to have over the whole sky within the timescale of a decade or so.

Although focused on the Euclid cosmology mission, the ideas and results from this paper may be of interest in many other applications that involve photo-z.

\section{Generation of the Photometric Catalogues}

In order to simulate the catalogues for this work, we use the COSMOS mock catalogues produced by \cite{kitzbichler&White2007}. The mocks are generated from semi-analytic galaxy formation models using galaxy merger trees derived from the Millennium N-body simulation. The corresponding cosmological parameters are $ \Omega_{m}=0.25$, $ \Omega_{\Lambda}=0.75$,$ \Omega_{b}=0.045$, $ h = 0.73$, $ n = 1$, and $ w_{\Lambda}=-1$. The mocks have an area of $ 1.4^{\circ}\times1.4^{\circ}$ and are magnitude limited at $ I_{AB} \leq 26$. For each galaxy the catalogues give the right ascension RA, declination DEC, the redshift z and the magnitudes for the filters Bj, g+, r+, i+, and Ks. The right ascension and declination are in the range $ \left[  -0.7^{\circ}, 0.7^{\circ}\right]$. There are a total of 24 mocks, each produced from a different wrapped cone that passes through the cubical simulation so that no object appears twice in a given cone. A given galaxy will however appear at different redshifts in the different mocks. Each mock catalogue contains approximately 600,000 objects, the majority of which are at $ z\leq 1.$ 

In order to produce a photometric catalogue with our own set of filters, we first identified an SED template, from the 10,000 available templates, that well matched the given $B_j$, $g+$, $r+$, $i+$, and $K_s$ photometry for each galaxy in the \cite{kitzbichler&White2007} catalogue, at its known redshift. This operation made use of a program kindly provided by Thomas Bschorr. These templates include a range of internal reddening, which ranges from $0 < A_{v} < 2$ magnitudes.  This chosen template was then used to compute ideal photometry (i.e. without any observational noise) for this galaxy in any other passband of interest. Intergalactic absorption in each template is compensated for using the Madau law \citep{Madau1995}. In order to match to the proposed Euclid weak lensing experiment, we consider only objects with $ I_{AB} \leq 24.5$.  Except for the cosmic variance analysis in the Appendix, we combine all 24 mocks and use a random sub sample to mimic a survey over a large area in the sky. Each of our simulations contains at least 100,000 objects.

\begin{figure}\label{graphics}

{\includegraphics[angle=0,width=9cm,height=6cm]{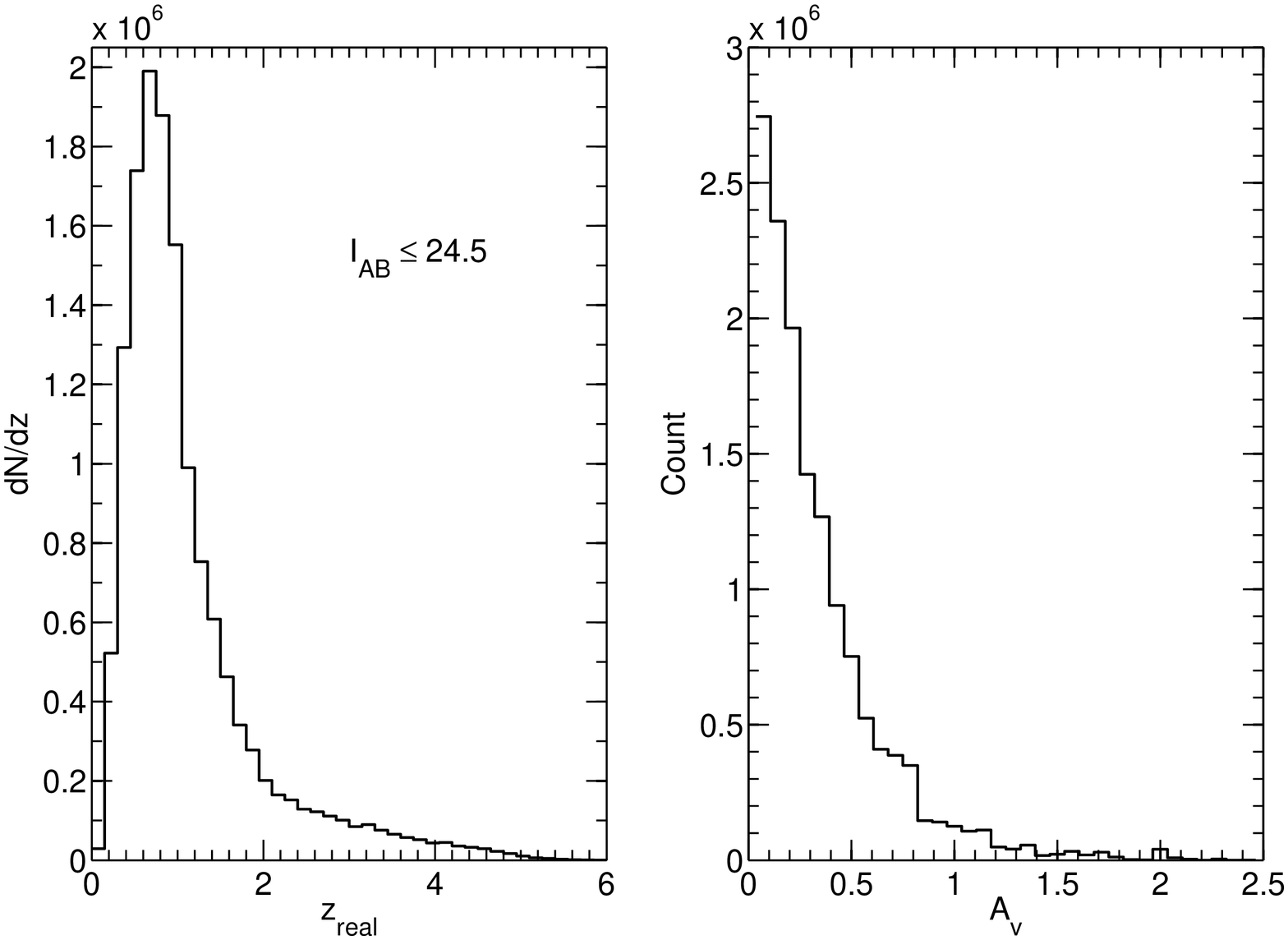}}

\caption{The left panel shows the $N(z)$ distribution of the artificial catalogue with $I_{AB} \leq 24.5$ used in this paper. The right panel shows the distribution of internal $A_V$ in the chosen SED templates for these galaxies.}

\label{mock_info}
\end{figure}

To generate a photometric catalogues as observed, this ideal photometric catalogue is degraded by adding Gaussian noise (in flux) to the photometry according to the three different sets of survey sensitivities that are listed in Table-1. All three sets contain the nominal near-infrared photometry expected from Euclid but have different choices for the depth in $grizy$, and therefore explore the requirements for the ground-based complement of the infrared survey.  Survey-A, which we generally find to be inadequate, uses the point source sensitivities from \cite{Cai2009}. This is therefore a possibly optimistic representation of a PanStarrs-1-like survey.  Survey-C, which we find exceeds the requirements, uses the PanStarrs-4 extended source sensitivities which is calculated by taking nominal PanStaar-4 point source sensitivities from \cite{Abdalla2008}, degraded by 0.6 mag to account for extended sources.  The proposed LSST large area survey would be expected to be slightly deeper than this. Finally, Survey-B is intermediate between these two, and approximates what could be expected from PanStarrs-2 or the DES.  The precise choice of these sensitivities is of course somewhat arbitrary, and they are adopted here for the sake of illustration.  Unless stated otherwise, all magnitudes stated in this paper are in the AB system. 

\begin{table}
 \begin{center}
    Assumed sensitivities for the representative surveys
  \begin{tabular}{crrr}
  \hline
  \hline
  Band &   Survey-A     &      Survey-B     & Survey-C      \\
   \hline
   \hline
g    & 24.66          & 25.53          &  26.10 \\
r    & 24.11          & 24.96          &  25.80 \\
i    & 24.00          & 24.80          &  25.60\\
z    & 22.98          & 23.54          &  24.10\\
y    & 21.52          & 22.01          &  22.50\\
\hline
& Euclid & NIR & \\
\hline
Y    & 24.00          & 24.00          &  24.00 \\
J    & 24.00          & 24.00          &  24.00\\
H/K    & 24.00          & 24.00          &  24.00\\
\hline
\end{tabular}
\caption{The filter sensitivities for different survey configurations considered. The values quoted here are $ 5\sigma$ errors in AB magnitude.}
\end{center}
\label{table-1}
\end{table}

\subsection{Estimating photo-z s using ZEBRA}

In this work the template-fitting photo-z code ZEBRA \citep{Feldmann2006} is used to produce photo-z for the galaxies in each of the observationally degraded catalogues. ZEBRA gives a single best fit redshift, which we call the ``maximum likelihood redshift'' and template type, together with their confidence limits estimated from constant $ \chi^{2}$ boundaries. ZEBRA also outputs the normalized likelihood functions $L(z)$ for individual galaxies in various formats, which we also use in this paper. $L(z)$ can be modified by a Bayesian prior, as desired, but is in any case normalised so that the integral over all redshifts is unity. Further information is available in the ZEBRA user manual.\footnote{http://www.exp-astro.phys.ethz.ch/ZEBRA/}.

\section{Performance on individual objects}

In this section, we compare the basic performance of the photo-z estimation by comparing the maximum likelihood photo-z with the known redshifts of the galaxies, for different choice of survey depths for the different simulations presented in Section 2. The aim of this section is to assess what sort of ground-based data is required to complement the Euclid infrared photometry and to develop techniques for the automated recognition and elimination of outliers with wildly discrepant photo-z. 

\subsection{Depth of ground-based photometry}

Using the photometric catalogues that were described in the previous section, and as degraded to simulate different survey configurations, we compare their photo-z performance.  We first bin the galaxies into narrow redshift bins on the basis of their ``observed'' photo-z.  We then use the bias(b) and the dispersion $\sigma_{z}(z)$ to parameterize the performance, defining these as follows:  The error per object ($ \delta z_{i}$) is 

\begin{equation}
\rm \delta z_{i} = \, (z_{\rm{real,i}}\,-\, z_{\rm{phot,i}})  
\end{equation}

where $ z_{\rm{real,i}}$ and $ z_{\rm{phot,i}}$ are the real and photometric redshifts of the ith galaxy. The mean bias in each photo-z bin $ \Delta_{z}(z)$ is then

\begin{equation}
\Delta_{z}(z)\,=\, \langle \delta z \rangle 
\end{equation}

The r.m.s. deviation in the photo-z estimation within the bin $ \sigma_{z}(z)$) is 
\begin{equation}
\sigma^{2} _{z}(z) \, = \, \langle ( \delta z_{i} - \Delta z)^{2} \rangle 
\end{equation}

and the total mean squared error (MSE) is given as

\begin{equation}
MSE(z)\, =\, \sigma^{2} _{z}(z) + \Delta^{2} _{z}(z)
\end{equation}

\begin{figure*}
\includegraphics[angle=0,width=8cm,height=7cm]{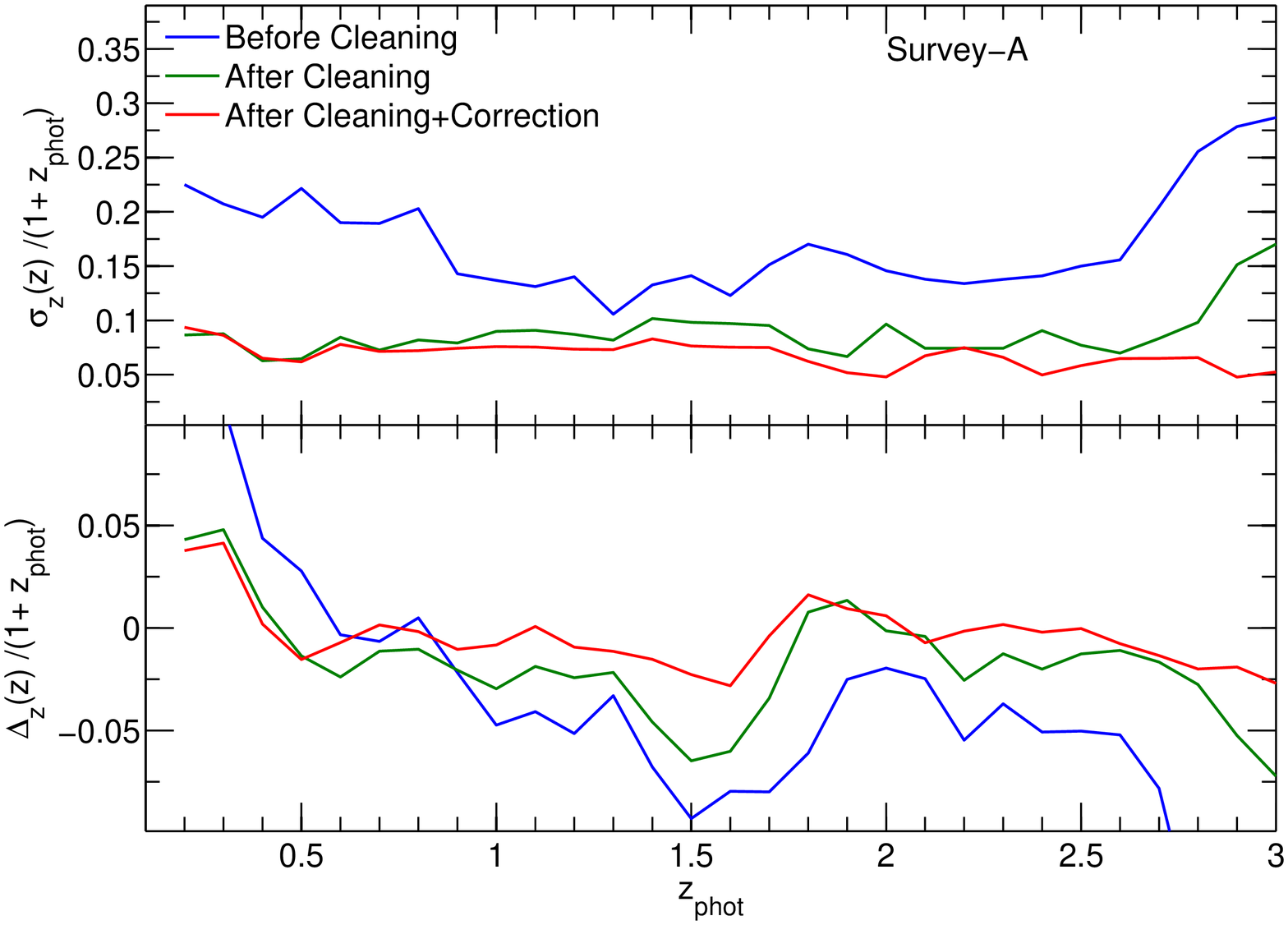}
\includegraphics[angle=0,width=8cm,height=7cm]{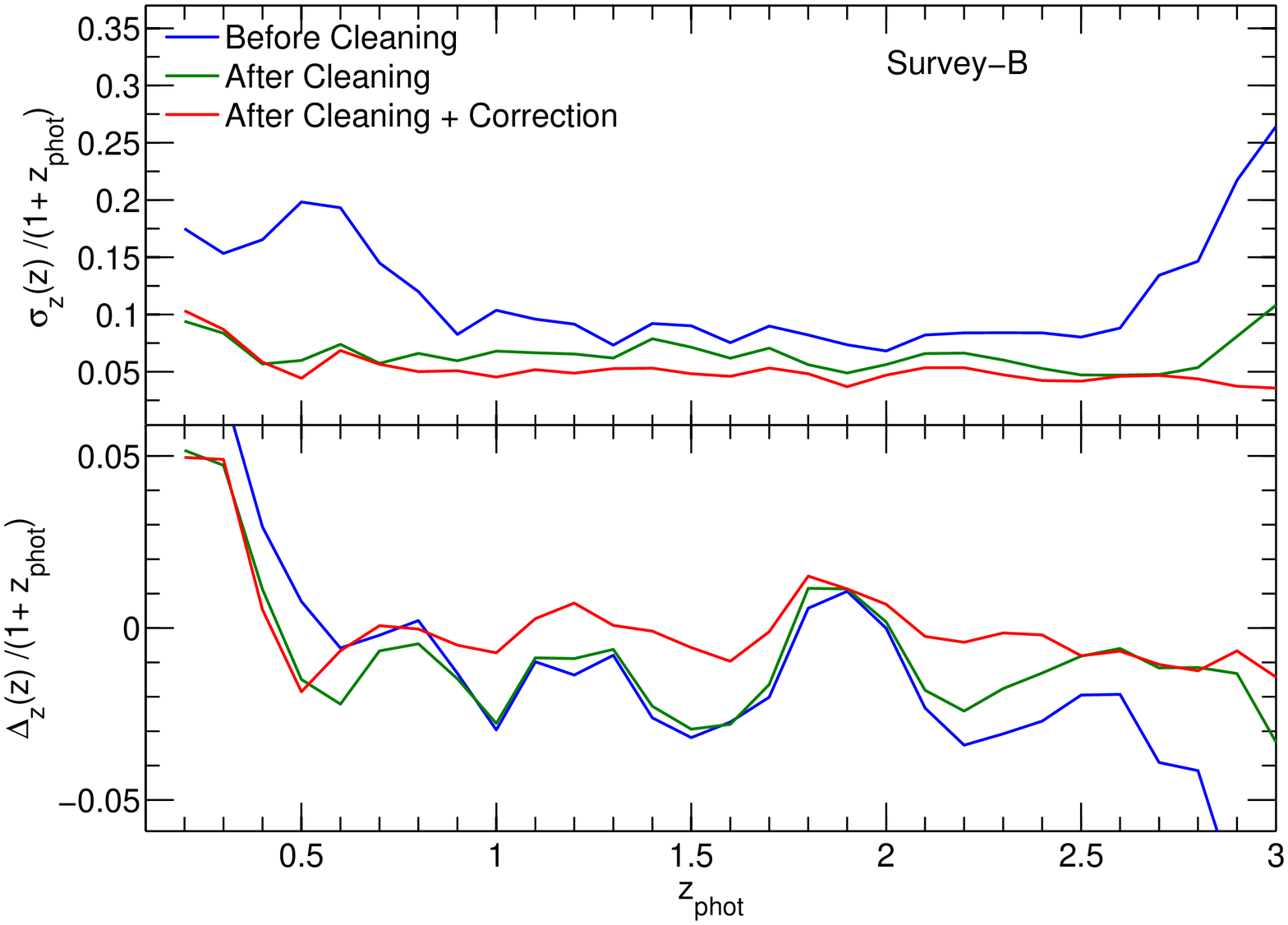}

\includegraphics[angle=0,width=8cm,height=7cm]{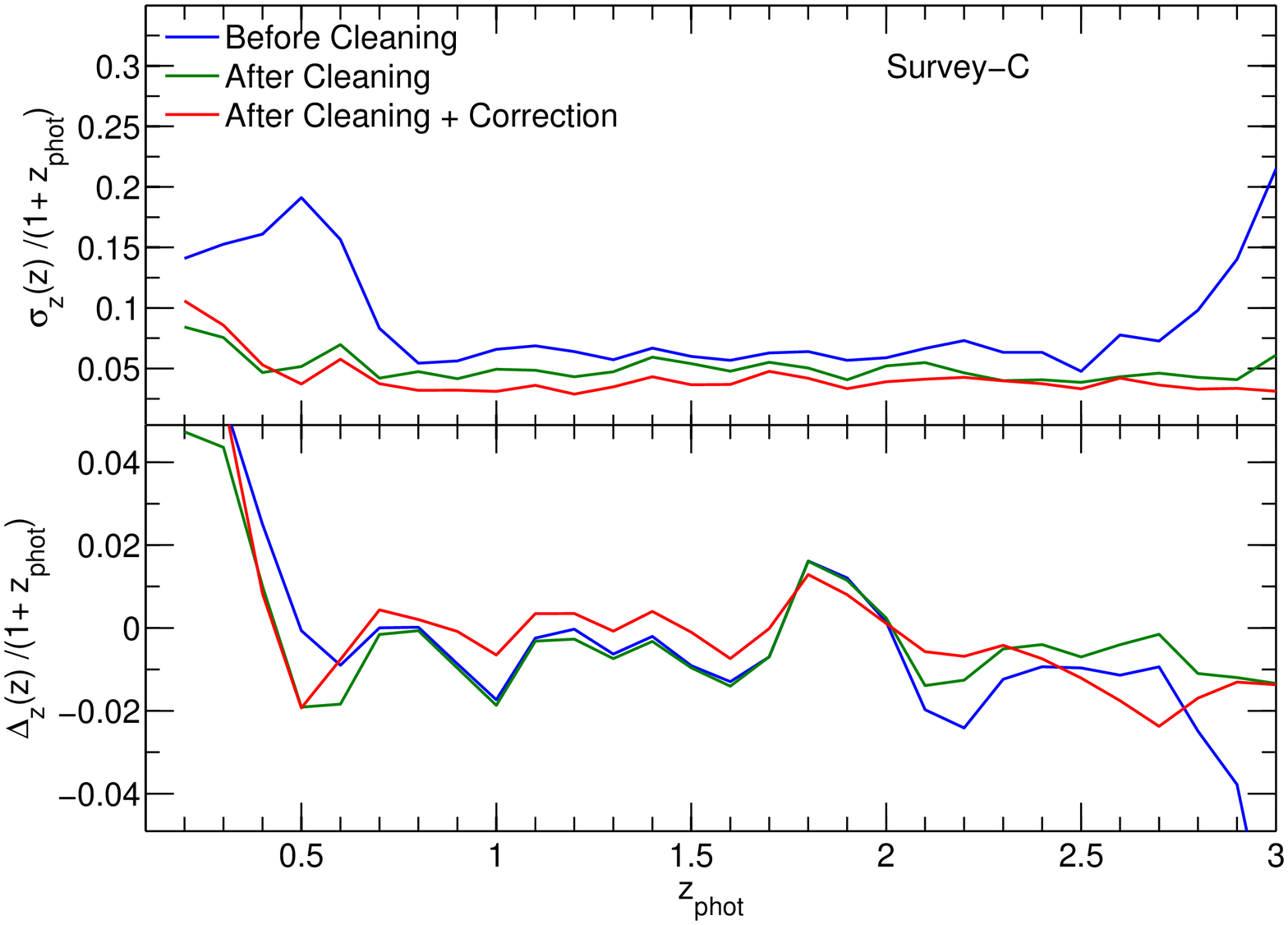}

\caption{ The overall performance of the survey-A, survey-B and survey-C whose dephts are as quoted in Table-1. The blue lines give the performance without cleaning and the green lines after cleaning and the red line gives performance after cleaning and applying correction. It is seen that with cleaning and correction $ \sigma_{z}(z)/(1+z)\, \leq$ 0.05 is almost reached in  all the cases and systematic bias is also reduced considerably. Survey-B after cleaning and correction reaches $ \sigma_{z}(z)/(1+z)\, \leq$ 0.05 easily. For survey-A 23$ \%$ for survey-B 13$ \%$ and for survey-C 9$ \%$ rejections were made after cleaning.}
\label{overall_performance}

\end{figure*}

In Figure-\ref{overall_performance} we show the $\sigma_{z}(z)$ and $\Delta_{z}(z)$ for the different survey configurations. The blue curves in all the panels give the initial performance of the photo-z code, without any attempt at removing outliers. As expected, increased depth in the optical ground photometry increases the reliability of the photo-z estimates. However, none of the configurations match, without cleaning, the requirement of $ \sigma_{z}(z)/(1+z)\, \leq$ 0.05, especially at the lower redshifts $ z \sim 0.5$ where many the galaxies in fact lie.  The green curves show the effect of removing outliers, recognized purely photometrically (see Section 3.2), and the red curve shows the effect of modifying the individual $L(z)$ as described in Section 3.3.

\subsection{A priori identification of outliers }

In template fitting, a likelihood function $L(z)$ is derived for each galaxy from which maximum likelihood photo-z is estimated. In an ideal case, with a well-defined photo-z estimate, the $L(z)$ has a single tight peak. Empirically, it is found that many galaxies with poor photo-z estimates have a bimodal likelihood distribution. We therefore developed an algorithm that searches for bimodality in the likelihood curves of each galaxy. If a likelihood function contains more than one peak separated by a certain pre-defined redshift difference and if the ratio between primary and secondary peaks is above a threshold value, then the galaxy is flagged as a likely outlier and can be rejected from the lensing analysis. This pre-defined threshold value can be tuned from simulations of the kind described here, or from spectroscopic measurements of actual redshifts. Of course, this procedure will undoubtedly remove some objects whose photo-zs are actually quite good, but the lensing analysis is stable to this kind of exclusion.

After removal of doubtful photo-z, the errors in $\sigma_{z}(z)$ and mean bias $\Delta_{z}(z)$ are dramatically reduced, as shown by the green lines in Figure-\ref{overall_performance}. The major improvement in $ \sigma_{z}(z)$ and $\Delta_{z}(z)$ come from rejection of catastrophic failures rather than a tightening of the ``good'' photo-z.  As the depth of the photometry increases, it is found that fewer objects need to be rejected to improve the photo-z estimates. In case of survey-A, we find that 23$ \%$  must be rejected to get below $ \sigma_{z}(z) \leq 0.05(1+z)$, for survey-B it is 12$ \%$ and for survey-C, only 9$ \%$. The trade off between beneficial cleaning and the wasteful loss of objects determines the robustness of the cleaning.  After the above cleaning has been performed, the fraction of $ 5\sigma$ outliers (catastrophic failures) is reduced below $ 0.25\%$ in all the three Surveys (see Table-2). It should be noted that we have not taken in to account priors such as the size or luminosity of the galaxies, which might further improve the performance.

\subsection{Modification of the likelihood functions}

We find that the photo-z estimates can be further improved by modifying the $L(z)$ on the basis of a relatively small number of spectroscopic redshifts, as follows: First we define a variable $P(z)$ for each galaxy obtained by integrating the $L(z)$. 

\begin{equation}
  P(z) = \int_{0}^{z} L(z') dz'
\end{equation}
For galaxies where we reliably know the real redshift $z_{real}$ (e.g. from spectra), we can compute $P_{real} = P(z_{real})$. If the likelihood functions of the galaxies have statistical validity, then the distribution of $P_{real}$ for all the galaxies for which one has spectra, i.e. $N(P_{real})$, must be uniform between the extreme $P$ values of 0 and 1, i.e.

\begin{equation}
N(P_{real}) dP_{real} = dP_{real}
\end{equation}

If this is found not to be the case, then one can argue that a modification or correction to the individual $L(z)$ is warranted. We approach this by constructing a global mapping between $P$ and $P'$ that is determined from all objects with reliably known redshifts, such that the distribution of $P'_{real}$ will be flat.  We can write  
\begin{equation}
\frac{dP}{dz} = \frac{dP}{dP'} \times \frac{dP'}{dz}
\label{dpdz}
\end{equation}

Note that 

\begin{equation}
\frac{dP'}{dP} \sim N(P_{real})
\end{equation}

and  $N(P') = 1$ when $P=P'$.  We then modify the L(z) for each galaxy, with or without a known redshift, to produce $L'(z)$ such that, for all z,

\begin{equation}
P'(z) = \int_{0}^{z} L'(z') dz'
\end{equation}

It is easy to see that this is given by the following simple correction to $L(z)$:

\begin{equation}
L'(z) = L(z) \times N(P(z)) 
\end{equation}

where N(P(z)) is the ``observed'' distribution of $N(P_{real})$ with applied mapping of $z$ to $P$ for each object using equation 11 above.

This procedure is clearly related to the application of a conventional Bayesian prior in redshift space, but is now applied in P (probability) space and is based on the absolute requirement of having a flat $N(P_{real})$ for meaningful likelihood functions.  The application of the global mapping of $P$ to $P'$ to all objects, independent of their nature, is of course arbitrary and cannot be rigourously justified. However, we find this approach works well, both here and later in the paper.

\begin{figure}\label{graphics}
{\includegraphics[angle=0,width=8cm,height=7cm]{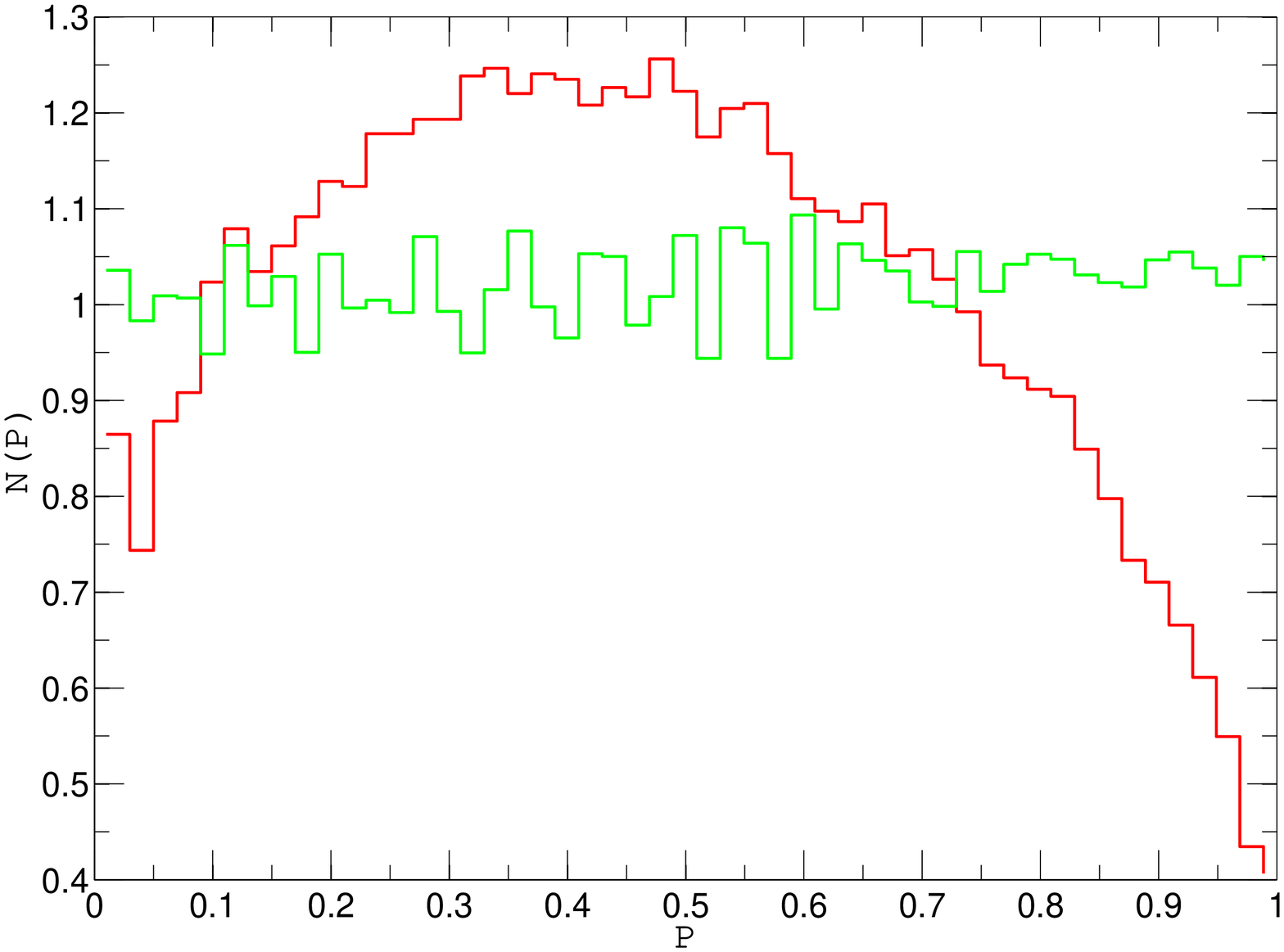}}

\caption{ Normalized N(P) before (red) and after (green) applying the correction scheme. In ideal case this distribution should be flat in P. }

\label{N(P)}
\end{figure}

This procedure is illustrated in Figure-\ref{N(P)}, in which the red line gives the $N(P_{real})$ for all the galaxies with Survey-C like sensitivities. We therefore compute an empirical correction so as to make the $N(P_{real})$ curve to be flat.  The green line in (Figure-\ref{N(P)} green line) is produced by using a sub sample of 1000 galaxies. Note that due to discrete binning in $P$ space there is noise introduced in the $ N(P(z))$ function and hence the corrected green line is noisy. This noise doesnot translate into noise in $z$ space.

From the peaks of the new likelihood function $ L'(z)$, we can also define a more accurate individual photo-z estimate as given by the red line in Figure-\ref{overall_performance}. We stress that it is not necessary to know the N(P) very accurately and an average correction of N(P) using a relatively modest number of spectra yields significant improvement to photo-z accuracy.

\begin{table}
 \begin{center}
    Percentage of $ 5 \sigma $ outliers ($ f_{cat}$) in 
  \begin{tabular}{crr}
  \hline
  \hline
  Survey &   Before Cleaning     &     After Cleaning      \\
   \hline
Survey-A    & 1.18          &  0.2300 \\
Survey-B    & 0.8820          &  0.2138 \\
Survey-C    & 0.8221          &  0.1776 \\

\hline
\end{tabular}
\caption{The percentage of $ 5 \sigma$ outliers in various surveys studied. $ f_{cat}$ reduces significantly once cleaning of the catalogue is performed, which identifies most of the outliers effectively. }
\end{center}
\label{table-2}
\end{table}

\begin{table*}
 \begin{center}
     $ \langle \frac{\sigma_{z}(z)}{1+z} \rangle$ for different surveys in the range $ 0.3 \leq z \leq 3.0$
  \begin{tabular}{crrr}
  \hline
  \hline
  Survey &   Before Cleaning     &     After Cleaning   & After Cleaning + Correction   \\
   \hline
Survey-A    & 0.1703          &  0.0884 & 0.0675 \\
Survey-B    & 0.1164          &  0.0640 & 0.0497  \\
Survey-C    & 0.0876          &  0.0492 & 0.0398 \\

\hline
\end{tabular}
\caption{ The $ \langle \frac{\sigma_{z}(z)}{1+z} \rangle$ for the three surveys studied. After cleaning and correction has been performed survey-B just about reaches $ \sigma_{z}(z)/(1+z) \sim 0.05$ Euclid requirements. }
\end{center}
\label{table-3}
\end{table*}

\begin{figure}\label{graphics}
{\includegraphics[scale=.35]{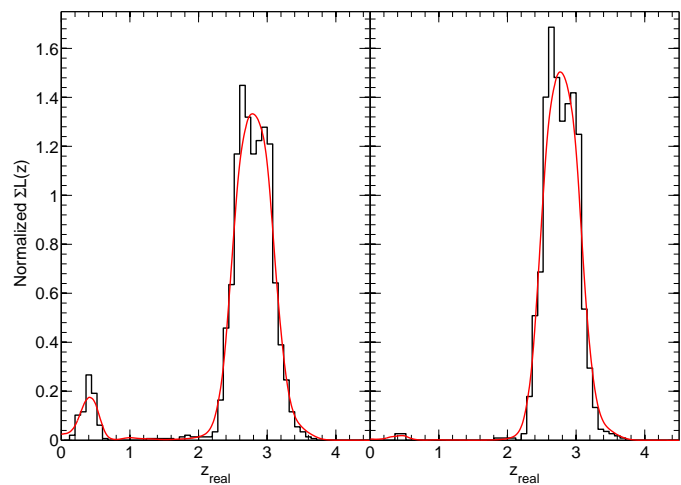}}

\caption{N(z) constructed from the $ \sum L(z)$ function before and after cleaning. Here the normalized histogram gives the real redshift distribution in the bin and line is the N(z) constructed from the $ \sum L(z)$ function. The left panel gives the redshift bin before cleaning and right panel gives after cleaning. The constructed N(z) clearly traces the catastrophic failures.}

\label{catastrophic_failure}
\end{figure}

\section{Characterization of N(z) from the likelihood functions }

\begin{figure}\label{graphics}
{\includegraphics[angle=0,width=8cm,height=7cm]{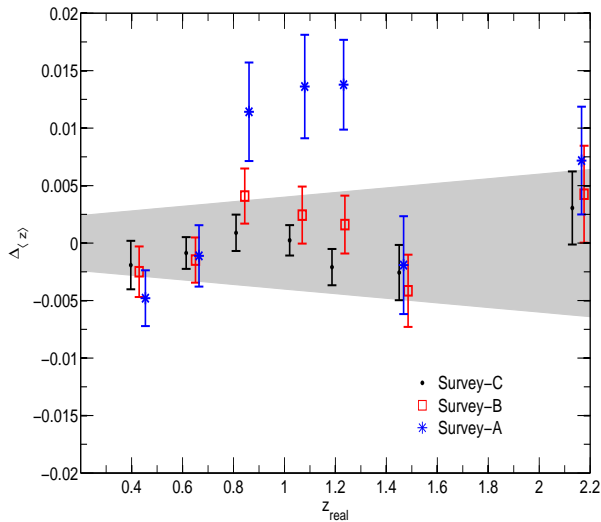}}

\caption{ The bias in the mean of the tomographic bins estimates from the Normalized $\sum L(z)$ functions for survey-C and survey-A  and survey-B. For survey-C, with cleaning for catastrophic failures and after applying correction gives $ |\Delta_{\langle z\rangle}/(1+z)| \leq 0.002 $. Here the shaded region is $ |\Delta_{\langle z\rangle}| = 0.002(1+z)$. We have introduced a small offset in x-axis values of survey-B and survey-C for legibility. }

\label{Deltaz_subplot}
\end{figure}

In weak lensing tomography the photo-z s are used to construct redshift bins which are then used to calculate the lensing power spectrum. The actual $N(z)$ of each bin must then be known for quantitative interpretation of the lensing signal. The mean of the distribution is most important parameter \citep{Amara&Refrigier2007} and we therefore focus on this.  Generally a single redshift estimator from the photo-z code (i.e. the maximum likelihood photo-z) is used to construct these bins.  However, if using these single redshifts, the $ \Delta_{\langle z\rangle}$ requirement cannot be reached, as clearly shown in Figure-\ref{overall_performance}. This is because the maximum likelihood redshifts cannot by construction trace the wings of the N(z) that lie outside of the nominal bins, or trace the remaining catastrophic failures associated with some of the photo-z. Therefore a more sophisticated approach is required.

As noted in the Introduction, one approach is to undertake a major spectroscopic survey of large numbers of representative objects in the bin and define the actual N(z) empirically in this way.  As discussed there, there are a number of practical difficulties of doing this.  

In this paper we explore a different approach, which is to characterize N(z) as the sum of the likelihood functions for each redshift bin. 
We define the mean redshift inferred from summing the likelihoods as:

\begin{equation}
\overline{z} = \langle \sum L(z)\rangle = \int_{0}^{\infty} z \sum L(z) dz 
\end{equation}
and the bias in estimating $ z_{\rm{real}}$ as
\begin{equation}
\Delta_{\langle z\rangle} = \rm{z_{\rm{real}}} - \overline{z} 
\end{equation}

We apply this approach using the same modification techniques described in Section 3.3.  The straight sum of the original likelihood functions is able to characterize the redshift distribution well, as seen in Figure-\ref{catastrophic_failure}, which shows for survey-C the summed $L(z)$ follows (visually) both the the catastrophic failures and the wings of the redshift bins well. If we apply the cleaning algorithm described above, the number of catastrophic failures are removed and wings are constrained more tightly. However, this approach alone is not in fact good enough to characterize the N(z) of the bins to the required precision of $ |\Delta_{\langle z\rangle}| \leq 0.002(1+z)$.

To characterize the bins more accurately, the $L(z)$ correction scheme as described in Section 3.3 was developed.  We compute $N(P)$ for each redshift bin separately, using a spectroscopically observed subsample of 800-1000 galaxies per bin.  After correction, the new likelihood functions $L'(z)$ for each galaxy, and therefore sometimes a new maximum likelihood redshift, is obtained. These are used to rebin the galaxies and the sum of the new $L'(z)$ are used to construct $N(z)$ for the bins. In Figure-\ref{Deltaz_subplot} the bias on the mean of the $N(z)$ is given for different redshift bins, and survey parameters.  The error-bars on each point shows the effect of randomly picking different subsets for the the spectroscopic calibration repeatedly.  In Figure-\ref{Deltaz_subplot} the shaded region gives the Euclid requirement of $ |\Delta_{\langle z\rangle}/(1+z)| \leq 0.002$ on the mean redshift of the redshift bins. The black dots are for survey-C, which easily reaches the Euclid requirements. The red open boxes are for survey-B and it just meets the Euclid requirement. The blue stars are for survey-A which do not meet the specifications as given by the shaded region. From this analysis we conclude that for a Euclid like survey, using a survey-B like ground based complement we can characterize the N(z) of the tomographic bins to a precision of $ |\Delta_{\langle z\rangle}/(1+z)| \leq 0.002$ and we need around 800-1000 random spectroscopic sub-sample per redshift bin to characterize them. 

The great advantage of this approach is that it sidesteps completely the problems associated with the presence of large scale structure in the spectroscopic survey fields, since the spectro-z are used to characterize, and globally modify, the photo-z estimates of individual galaxies, and not to characterise directly the N(z), which will clearly be affected by such structure. It is also less susceptible to incompleteness in the measurement of spectroscopic redshifts, either in selection for spectroscopic observation or in success in measuring a redshift. That said, it is based on an assumption that, for a given spectroscopic target at some maximum likelihood photo-z, the ability to measure a redshift will not systematically depend on the location of the real redshift in P-space.
\begin{figure}\label{graphics}
{\includegraphics[angle=0,width=8.5cm,height=8cm]{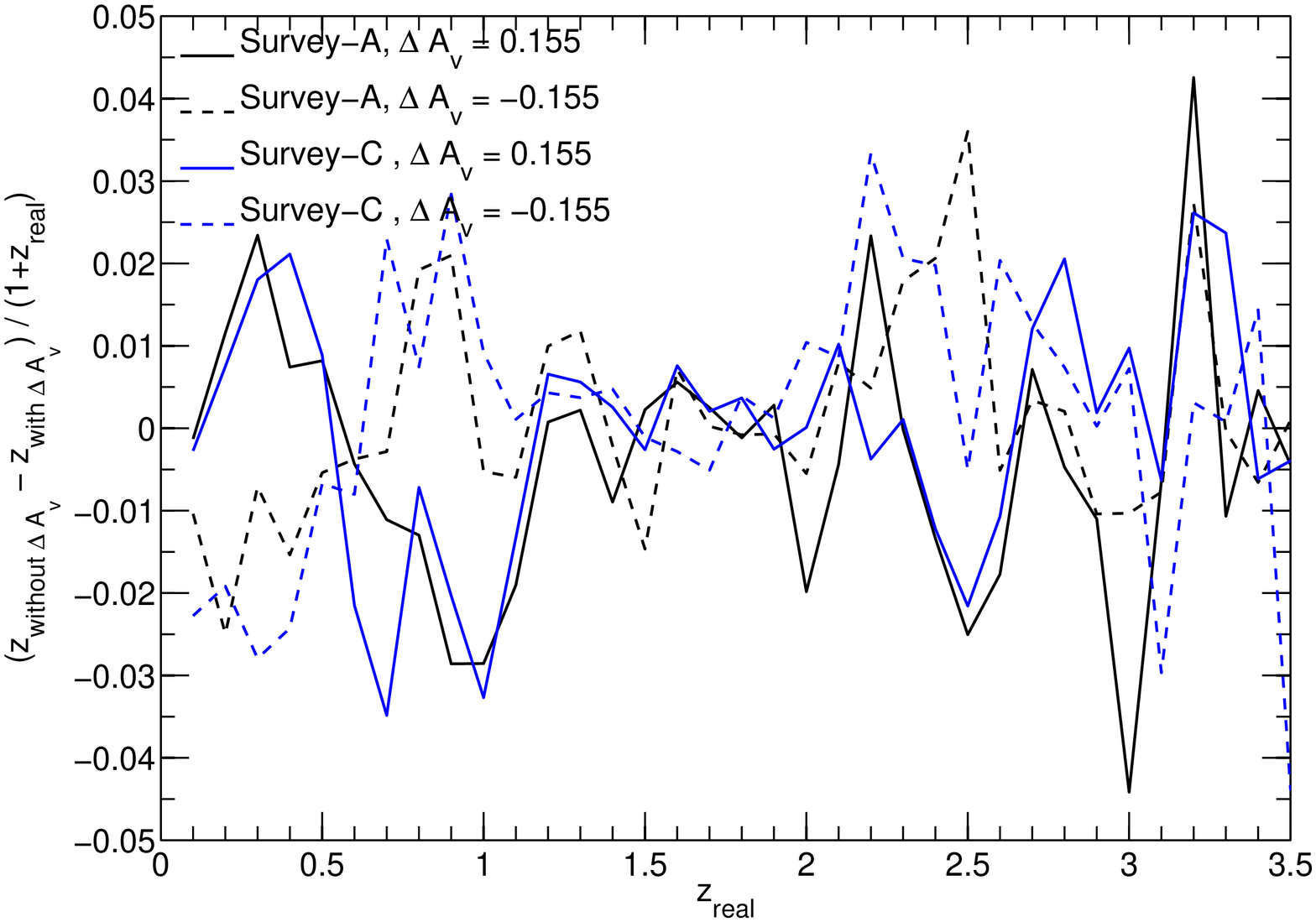}}

\caption{The bias introduced in photo-z estimation due to small offset in photometry after an average correction for effects of reddening. For survey-A and survey-C simulations two opposite $ \Delta A_{v}$ offsets were investigated 0.115 and -0.115. Changing the sign of $ \Delta A_{v}$ leads to a more or less mirror inversion in the bias.}

\label{extinction_bias}
\end{figure}

\begin{figure}\label{graphics}
{\includegraphics[angle=0,width=9cm,height=6cm]{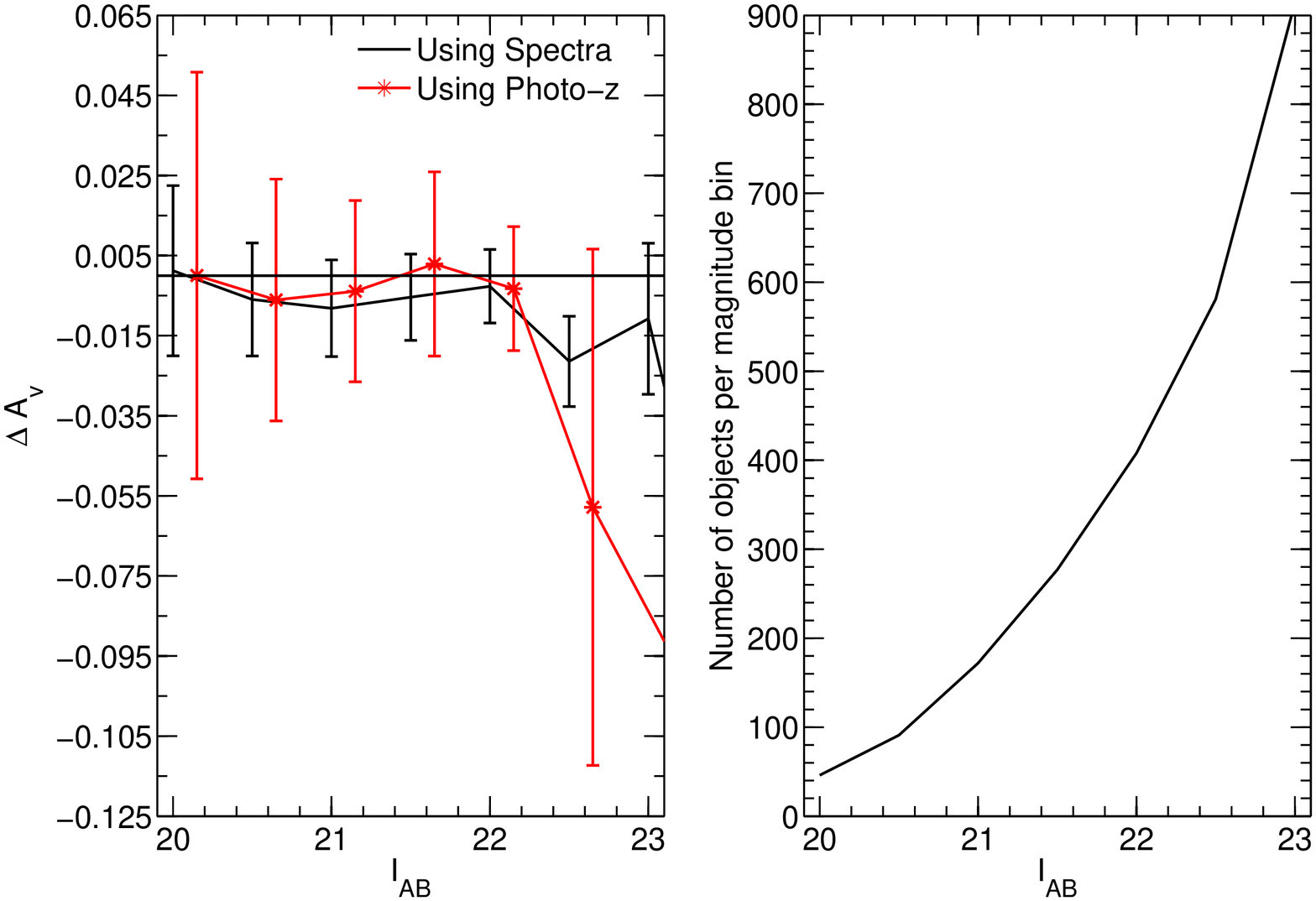}}

\caption{The left panel gives estimates of $ \Delta A_{v}$ for different magnitude bins (in the configuration of Survey-C). The red line is obtained using internally computed photo-z, without knowledge of the redshifts of the galaxies, and the blue line is using spectroscopic information. Here we have introduced a small offset in x- axis  on the red line for legibility. The right panel is the number of objects per magnitude bin. With spectra, at $ I_{AB} \sim 22$ magnitude bin around 350 spectra are sufficient to estimate $ \Delta A_{v}$ accurately.  }

\label{simultaneous_fit}
\end{figure}

\begin{figure}\label{graphics}
{\includegraphics[angle=0,width=8cm,height=6cm]{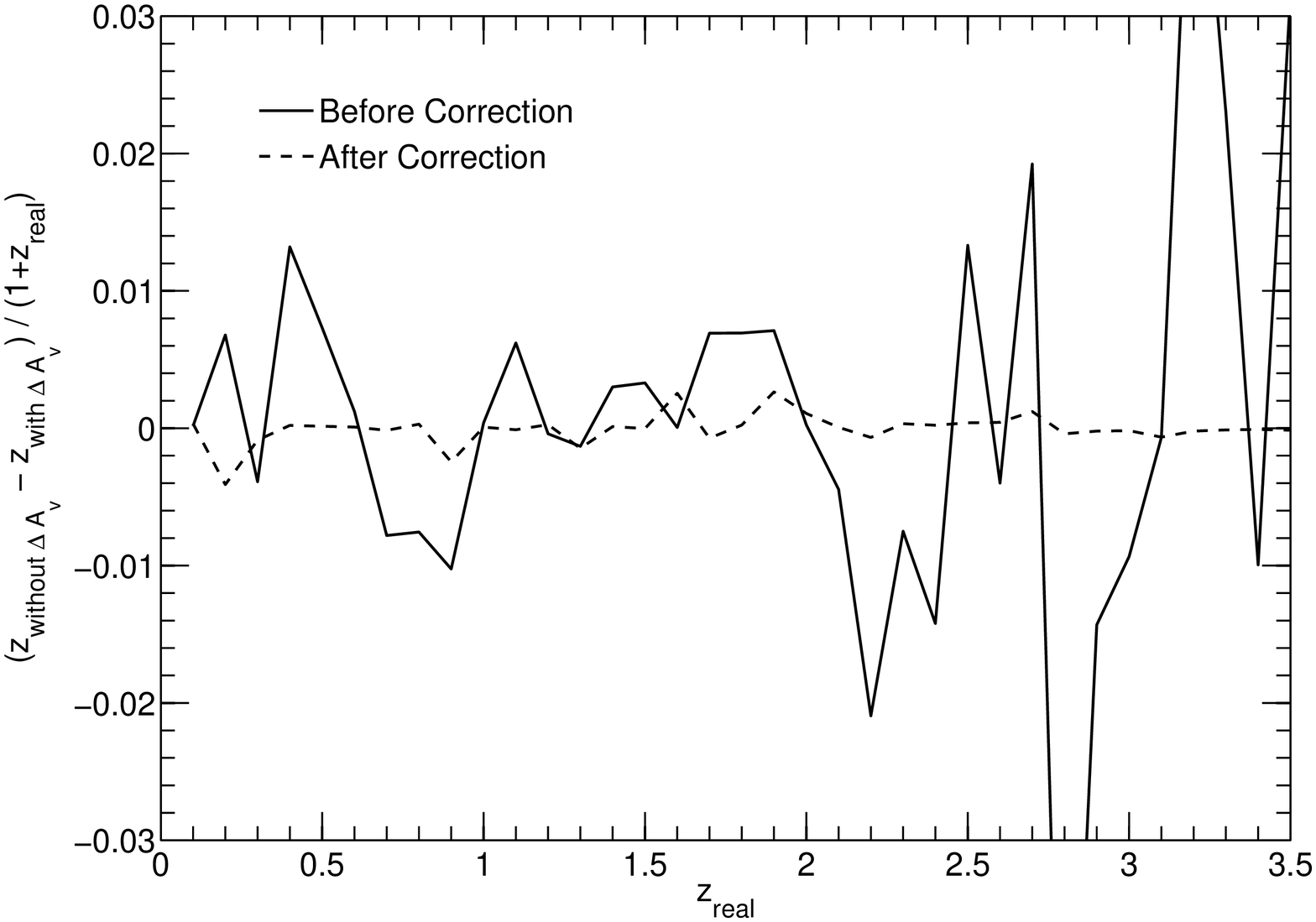}}

\caption{Performance of the photo-z estimates before and after estimated $ \Delta A_{v}$ correction have been applied. using this correction scheme the systematic bias introduced due to $ \Delta A_{v} $ is significantly reduced below 0.002(1+z) level. }

\label{simultaneous_fit_performance}
\end{figure}

\section{Internal Calibration of Galactic Foreground Extinction}

In this section we explore the effect that errors or uncertainties in foreground Galactic extinction can have on photo-z, and examine whether the photometry of large numbers of galaxies, with or without known spectroscopic redshifts, can be used to determine an improved extinction map and locally correct the extinction. This latter aspect is an extension of the iterative adjustment of photometric zero-points that is now standard in many template-fitting photo-z codes.  

Extragalactic photometry is routinely corrected for the effects of foreground Galactic extinction using reddening maps and an assumed extinction curve. In practical terms, the effect that we should therefore worry about is an error or uncertainty in the $A_V$, i.e. a $\Delta A_{v}$, which may be positive or negative.  This will cause galaxies to be either too red, or too blue, in the photometric input catalogue.

In this section we construct a catalogue containing $10^{4}$ objects down to $I_{AB} \sim 24.5$.  This mimics a roughly 0.1 deg$^2$ region of the Euclid survey.  We consider photometry with the accuracy expected from both survey-A and survey-C, and then perturb these catalogues by applying a standard mean reddening law \citep{Cardelli1989} with a relatively large $ \Delta A_{v} \sim 0.155$, in both the positive and negative directions. We then compare the photo-z for the galaxies with and without these $\Delta A_V$ offsets. Figure-\ref{extinction_bias} show the bias between these photo-z estimates as functions of redshift.  The bias fluctuates in redshift in a somewhat random way, with large systematic excursions at low redshifts.  Interestingly, but perhaps not surprisingly, changing the sign of the $\Delta A_V$ leads to a largely mirror effect on the redshifts, suggesting that the response of the photo-z scheme to errors in $A_V$ is linear.

Although we chose quite a large ``worst-case'' error in $A_V$, the biases with redshift seen in Figure-\ref{extinction_bias} are almost ten times worse than the $0.002(1+z)$ photo-z bias that can be tolerated by Euclid's precision cosmology. Large-scale redshift-dependent biases in the photo-z are particularly worrisome as they mimic the effect of cosmological parameters.  We therefore explore the possibility of iteratively identifying the residual $\Delta A_V$ error as follows.  We assume that we will know the wavelength dependence of the reddening in a given field and that the problem is therefore in determining the $A_V$.  At high galactic latitude, we expect that the wavelength dependence could be determined from very large areas of sky, but that $A_V(b,l)$ may vary on small scales.  

To estimate the ability of the photometry to determine $\Delta A_V$, we take the input photometric catalogue and ``correct'' it for a wide range of assumed $\Delta A_V$ around zero.  We then run ZEBRA on each of these corrected catalogues and take the $\sum \chi^{2}_{min}$ of the all the individual galaxies (i.e. the sum of the ``best fit'' chi-squareds). The value of $ \Delta A_{v}$ that produces the minimum $ \sum \chi^{2}_{min}$ is taken as the best estimate of $ \Delta A_{v}$ in that region.  The exercise can be undertaken with the redshifts of the galaxies as a free-parameter, or by assuming that the galaxies have known redshifts, and looking at the $\sum \chi^{2}_{min}$ amongst the templates at the known redshifts for each galaxy. The sample used here is magnitude limited to $I_{AB}  \leq 24.5$.

To obtain an error bar on $\Delta A_V$ we compute a reduced chi squared ($ \chi^{2}_{r}$).  For this we need to know the total degrees of freedom available in the template fitting approach, which is non-trivial since it is unclear how many degrees of freedom are associated with the 10,000 templates.  We assess this by requiring that the $ \chi^{2}_{r}$ be unity (using Survey-C, although this should not be important) and find that this gives a dimensionality close to 3, which sounds reasonable given what is known about galaxy spectra (see \cite{1995AJ....110.1071C}). A rule of thumb for estimating the uncertainty in one parameter gives \citep{1976ApJ...210..642A}.
\begin{equation}
\chi^{2}_{1\sigma\, \rm{confidence\, level}} = \chi^{2}_{min} + 1
\end{equation}
Hence the $ 1\sigma$ uncertainty in $ \Delta A_{v}$ estimate is given by the values of $ \Delta A_{v}$ which are below $ \chi^{2}_{r} + 1/DOF$ values.

To estimate the effect of photometric noise in estimating $\Delta A_{v}$ in this internal way, we consider objects in magnitude bins in $ I_{AB}$.  The results are shown in Figure-\ref{simultaneous_fit}.  We find that it is worth using only galaxies with relatively high S/N photometry, i.e. with the adopted survey parameters, down to $I_{AB} \sim 22$.  Below this level, the estimate degrades appreciably. The addition of spectroscopic redshift reduces the error bar significantly, but the method is still practicable down to the same magnitude limit and yields an error on $\Delta A_V$ of order 0.01 (with known redshifts) or 0.02 (without).  To close the loop, we show in Figure-\ref{simultaneous_fit_performance} the bias in photo-z introduced by applying 0.01 error to $\Delta A_V$ which shows improvement in photo-z errors by a factor of 10.

\section{Impact on photo-z from blended objects at different redshifts}

Sometimes multiple galaxies will overlap on the sky, and the photometry will be a composite of the two spectral energy distributions. Even with spectroscopy, such objects have composite spectra rendering their spectroscopic redshift estimation non-trivial. In this final section we explore the effects of this blending on photo-z estimates. We simulate many such blended objects by constructing composite spectral energy distributions constructed from galaxies at different redshifts, different colours and a wide range of relative brightnesses, from dominance of one through to dominance of the other. For definiteness we look at the photo-z behaviour for a survey-C like survey.

We select several objects from the main COSMOS mock catalogues at different redshifts and having different colours and normalize their fluxes to have the same $ I_{AB}$ brightness.  We then adjust the brightnesses of the two objects by $\pm 6$ magnitudes relative to each other, produce a co-added spectrum by averaging the fluxes, and then renormalise the resulting composite back to have $ I_{AB} = 23.5$, i.e. one magnitude above the survey limit.  Gaussian noise is then added in the usual way to the composite SED to represent the Survey-C sensitivities.

\begin{figure*}\label{graphics}

\scalebox{.85}{\includegraphics*[.1in,3.1in][8.1in,9.in]{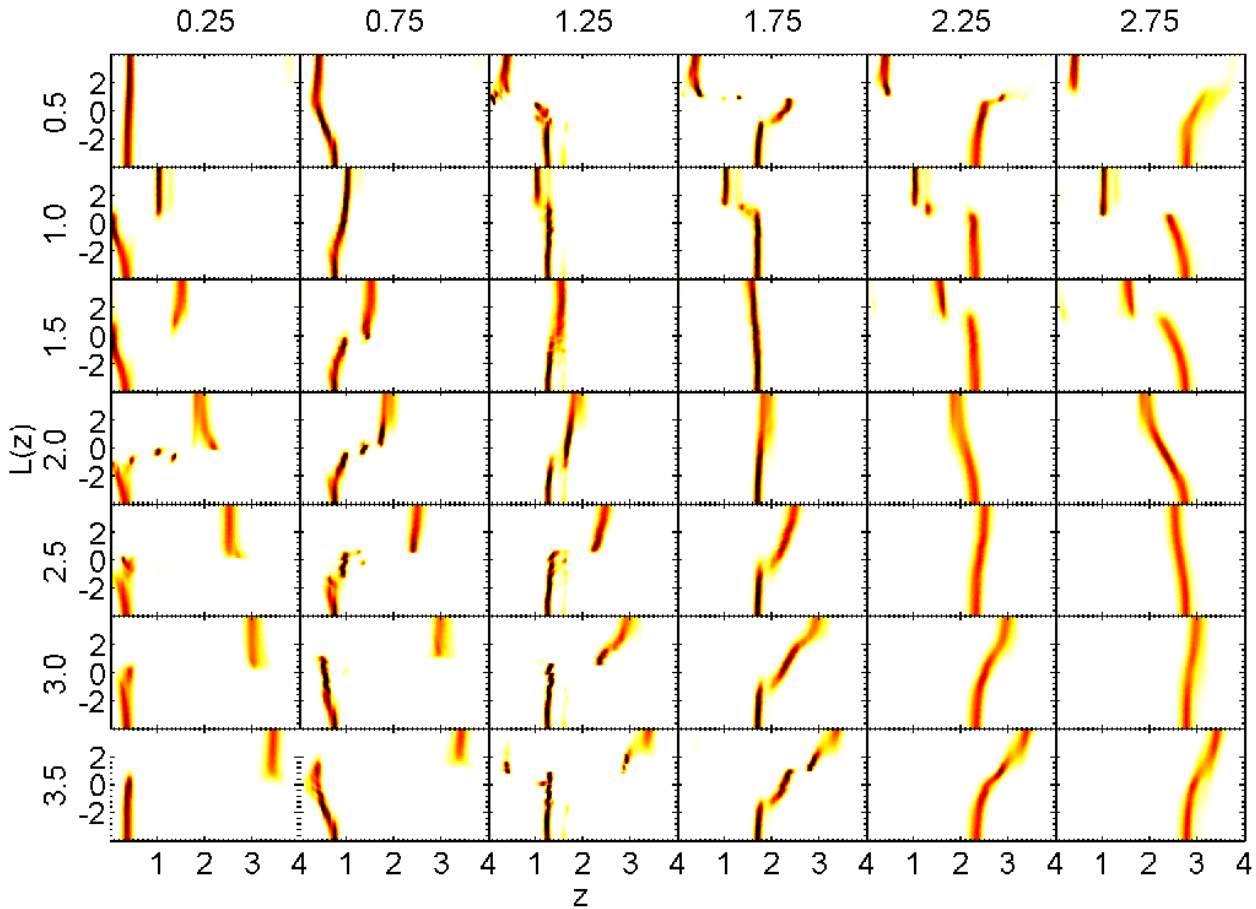}}
\caption{ L(z) functions of 42 pairs of blended objects. The objects were boosted to [-6 6] magnitude differences in the I band and variation of likelihood functions are observed. Here when the low redshift object is brighter than the high redshift one and the likelihood function traces low redshift function. When both the objects are equally bright we see the bimodality in the likelihood functions and as magnitude difference increases the likelihood function jumps to trace the high redshift object. The degeneracy of the likelihood functions at $ \Delta m \sim 0$ means that the redshift estimation gets completely unreliable at that region. Here the objects are taken for survey-C like depth. The objects are chosen such that in each pair one object is red and the other is a blue galaxy.}
\label{mergedobject}
\end{figure*}

ZEBRA is then run on these set of blended objects and the resulting likelihood curves of each composite object, along with a single maximum likelihood redshift, are output. In Figure-\ref{mergedobject}, each box represents a single composite object consisting of a red galaxy (whose redshift is indicated on the left) and a blue galaxy (whose redshift is indicated along the top).  In each panel, the likelihood curve is plotted as a function (vertically) of the adopted magnitude difference. 

We see from Figure-\ref{mergedobject} that if a pair of objects consist of two galaxies at similar redshifts then there is a smooth transition from one redshift to the other. When the magnitude difference is 2 magnitudes, or more, the redshift of the brighter galaxy is returned. When the contrast is lower, the maximum likelihood redshift transitions smoothly between the two, but will not accurately represent either component. When the pairs are at different redshifts i.e. $ \Delta z \geq 0.75$ the returned redshifts still trace the redshift of the brighter of the pair for $ \Delta I_{AB} > 2$ magnitudes. However, for the region $ \Delta I_{AB} < 2$, the behaviour is more varied. Bimodality in the $L(z)$ is often seen, and sometimes a local maximum at a third intermediate redshift. There are often sharp transitions. For the classically degenerate redshift pairs i.e. [{.25,.5},{2.5,3.5}], the photo-z generally make a sharp transition between the two redshifts. 

Our general conclusion from this analysis is that the photo-z of blended objects are trustworthy only if the second component is at least two magnitudes fainter than the primary (in a waveband close to the middle of the spectral range of the photometry). At smaller magnitude differences the photo-z can be corrupted in a way that is not always recognizable, and we suggest therefore that these blended objects be recognized morphologically from the images, and excluded from the analysis.  Fortunately, one would probably want to do this anyway for a lensing analysis because their shape measurements would be hard to use quantitatively.

\section{Conclusions}

In this work we have investigated a number of issues that could potentially limit the photo-z performance of deep all-sky surveys, and thereby impede the ambitious precision cosmology goals of survey programs such as the proposed ESA Euclid mission.  In each case, we find that, while standard techniques do not get to the required accuracy, simple new approaches can be developed that appear to get to the required performance.

Knowledge of the redshifts enters into weak-lensing analyses at two distinct steps: first, the selection of objects for shape cross-correlation, and second the accurate knowledge of the mean redshifts of these objects. For practical reasons, the first step will likely require the use of photo-z for the foreseeable future.  A major motivation for the current work has been to develop techniques that rely on photo-z also for the second step, bypassing the need for very large and highly statistically complete spectroscopic surveys (e.g. \citep{Abdalla2008}). This approach thereby avoids the substantial practical difficulties that will be encountered in such spectroscopic surveys from incompleteness and the effects of large scale structure. A separate Appendix explores the latter effects in some detail.

The work is based entirely on simulated photometric catalogues that have been constructed to match the expected performance of three generic ground-based surveys, combined with the expected near-IR imaging photometry from Euclid.  To construct these catalogues, we use the same set of 10,000 templates as used for the template-fitting photo-z program (ZEBRA). This possibly circular approach allows us to remove the choice of templates as a variable.  We believe it is justified at the current time by the very impressive performance of template-fitting photo-z codes applied to the deep multi-band COSMOS photometry, which strongly suggest that the choice of templates will not be a limiting factor at the required level, although further refinement will be desirable.

The analysis is conveniently summarized in terms of the two main requirements on photo-z for precision weak lensing, namely to obtain an r.m.s. precision per object of $\leq 0.05(1+z)$ and a systematic bias on the mean redshift of a given set of galaxies of $\leq 0.002(1+z)$.
Our main conclusions may be summarized as follows:

\begin{center}
\begin{itemize}
\item To achieve an r.m.s. photo-z accuracy of $ \sigma_{z}(z)/(1+z) \sim 0.05 $ down to $I_{AB} \leq 24.5$, we need the combination of ground-based photometry with the characteristics of Survey-B (similar to PanStarrs-2 or DES) and the deep all-sky near-infrared survey from Euclid itself.  This performance also requires the implementation of an ``a priori'' rejection scheme (i.e. based on the photometry alone, without knowledge of the actual redshifts of any galaxies) that rejects 13$ \%$ of the galaxies and reduces the fraction of $ 5\sigma$ outliers to below $ f_{cat} < 0.25\%$. There is a trade off between the rejection of outliers and the loss of ``innocent'' galaxies with usable photo-z. Deeper photometry improves both the statistical accuracy of the photo-z and reduces the wastage in eliminating the catastrophic failures, and the combination of survey-C with Euclid near-infrared photometry achieves $ \sigma_{z}(z)/(1+z) \sim 0.04 $ after 9$ \%$ rejection.

\item A good way to determine the actual $N(z)$ of a set of galaxies in a given photo-z selected redshift bin is to sum their individual likelihood functions.  We find that the $\sum L(z)$ function represents well both the wings of the $N(z)$ and the remaining catastrophic failures. outliers.  However, to reach the required performance on the mean of the redshifts $| \Delta_{\langle z\rangle} | \leq 0.002(1+z)$ with the Survey-B, or deeper survey-C, combination (together with Euclid infrared photometry), we had to implement a modification scheme on the individual $L(z)$. This is based on the spectroscopic measurement of redshifts for a rather small number of galaxies (less than 1000) with relaxed requirements on statistical completeness (and no dependence on Large Scale Structure in the spectroscopic survey fields). We then require that the distribution of the actual redshifts within the probability space that is defined by the individual $L(z)$ should be flat across the sample as a whole.  This should be true of any set of galaxies, leading to relaxed requirements on sampling of the redshift survey.  This approach is similar to the application of a Bayesian prior on the redshifts, but is performed in probability space.  Although it cannot be rigourously justified, it is found to work well in practice.

\item  We find that uncertainties in foreground Galactic reddening can have a serious effect in perturbing the photo-z, with a net sign that varies erratically with redshift. However, we also find that such errors in $A_{v}$ can be identified internally from the photometric data of galaxies, either with or without spectroscopic redshifts. This procedure works best for galaxies with relatively high S/N photometry $ I_{AB} \leq 22$. The required number of galaxies suggests that a reddening map on the scales of 0.1 deg$^{2}$ can be internally constructed from the data on galaxies without known redshifts, or from a few hundred galaxies with spectroscopic redshifts.

\item We also explored the effect of blended objects. The photo-z of the composite spectral energy distribution is a good representation of the redshift of the brighter object as long as the magnitude difference is large, i.e. $ \Delta I_{AB} > 2.$ When the galaxies are more similar in brightness, $ \Delta I_{AB} < 2$ there is a wide-range of behaviour. In some cases, multi-modal likelihood functions appear, while in others there is a sharp transition from one redshift to another, sometimes with a local maximum at a third, completely spurious redshift.  In still others, the likelihood function smoothly transitions between the two redshifts with a single peak at an intermediate redshift.  Our conclusion is that composite objects with $ \Delta I_{AB} < 2$ should be recognized morphologically from imaging data and removed from the photo-z analysis.

\end{itemize}
\end{center}

The general conclusion of our study is that while reaching the photo-z performance required for weak-lensing surveys such as Euclid will not be trivial, the implementation of new techniques, coupled with internal calibration of e.g. foreground reddening from the photometric data itself, will allow the required performance to be attained.  If more reliance is placed on the photo-z themselves, then this leads to a significant simplification of the otherwise challenging requirement for spectroscopic calibration of large scale photometric surveys.

\section*{Acknowledgements}

We thank Thomas Bschorr for the use of his template matching program, and Pascal Oesch and Robert Feldmann for help in running ZEBRA.  We have benefitted during the course of this work from useful discussions with many members of the Euclid Science Study Team (ESST) and the Euclid Imaging Consortium (EIC). 

\bibliographystyle{mn} 
\bibliography{mybibliography}

\appendix
\section{Sampling requirements for direct spectroscopic calibration}

In this Appendix, we analyse the number of independent fields that are required if one takes the approach of determining N(z) for a given redshift bin from spectroscopic observations of a representative set of galaxies from that bin.  The requirement derives from desiring that the effects of large scale structure in the galaxy distribution, also known as cosmic variance, are at most equal to the Poisson noise in determining the error in $\langle z\rangle$.

For the sake of definiteness, we use the same set of 24 COSMOS mock catalogues \citep{kitzbichler&White2007} as used for the main paper. These are all derived from the same numerical simulation (the ``Millenium Run") but the 24 light cones sample this in such a way that the large scale structure at any given redshift is independent from one mock to the next.  These different light cones therefore give a measure of the variance that is introduced by large scale structure.

We compute the effects of this large scale structure on $ \langle z\rangle$, the average redshift of a set of galaxies in some defined redshift bin as follows.  We first look at all the galaxies in this redshift bin, across the whole 2 deg$^{2}$ field and across all of the mocks. The variance of their individual redshifts is given by the following expression,

\begin{equation}
\sigma^{2} = \frac{1}{n} \sum(z-\overline{z})^{2} = \frac{1}{n} (\sum z)^{2} - (\overline{z})^2
\end{equation}

For a top-hat distribution of redshifts within a redshift bin of width $\Delta z$, this would be given by 

\begin{equation}
\sigma^{2} \approx \frac{(\Delta z)^{2}}{12}
\end{equation}

We then imagine carrying out a redshift survey using a spectrograph with a particular field of view, which is assumed to be square.  We presume that galaxies (within this redshift bin) are randomly selected spatially within this field of view. We then compute for each mock $i$, the average redshift of the $n_{s,i}$ galaxies that fall within the spectrograph field of view.  Each field (or mock) yields an estimate of the underlying $\langle z_{i}\rangle$, which we denote by $ \zeta_{i}$. The variance of this quantity across the $m$ mocks is then computed as

\begin{equation}
\sigma_{\zeta}^{2} = \frac{1}{m}\sum(\zeta-\overline{\zeta})^{2} = \frac{1}{m}(\sum \zeta)^{2} - (\overline{\zeta})^{2}
\end{equation}

In the Poisson case, this variance would be equal to the variance coming from the $n_{s}$ galaxies in the field, which will be approximately given by $\sigma^{2}/n_{s}$, or more precisely by

\begin{equation}
\sigma_{P}^{2} = \frac{\sigma^{2}}{m} \sum_{i=1}^m \frac{1}{n_{s,i}} = \sigma^{2} \langle \frac{1}{n_{s,i}}\rangle
\end{equation}

For spectrograph fields of view that are much smaller than the mock, we can repeat the exercise for different locations of the spectrograph field within the 2 deg$^{2}$ field of the mocks, averaging the calculations of $\sigma_{\zeta}^{2}$.  We consider spectrograph fields of view given by $1.4/N$ deg, with $N = 1,6$, yielding in each case $N^2$ different locations with the 2 deg$^2$ field.

The difference between the ``observed'' $\sigma_{\zeta}^{2}$ and that from the Poisson variance $\sigma_{P}^{2}$ is the contribution of the large scale structure, or cosmic variance, $\sigma_{CV}^{2}$.

\begin{equation}
\sigma_{CV}^{2} = \sigma_{\zeta}^{2} - \sigma^{2}\langle\frac{1}{n_{s,i}}\rangle
\end{equation}

Given these estimates, we can establish, for each redshift bin and for each spectrograph field of view, a critical number of spectroscopic redshift measurements at which the cosmic variance will be equal to the Poisson variance.

\begin{equation}
N_{crit} = \frac{\sigma^{2}}{\sigma_{CV}^{2}}
\end{equation}

As the number of spectroscopic redshifts reaches this level, the standard deviation in the average redshift estimate $\zeta$ is already root-2 times higher than would be the case for Poisson noise alone, and obtaining more spectra in a given field will lead to little improvement in the estimate of the global mean redshift $\langle z\rangle$. 

\begin{figure*}\label{graphics}

\scalebox{0.75}{\includegraphics*[1.in,4.8in][7.8in,9.6in]{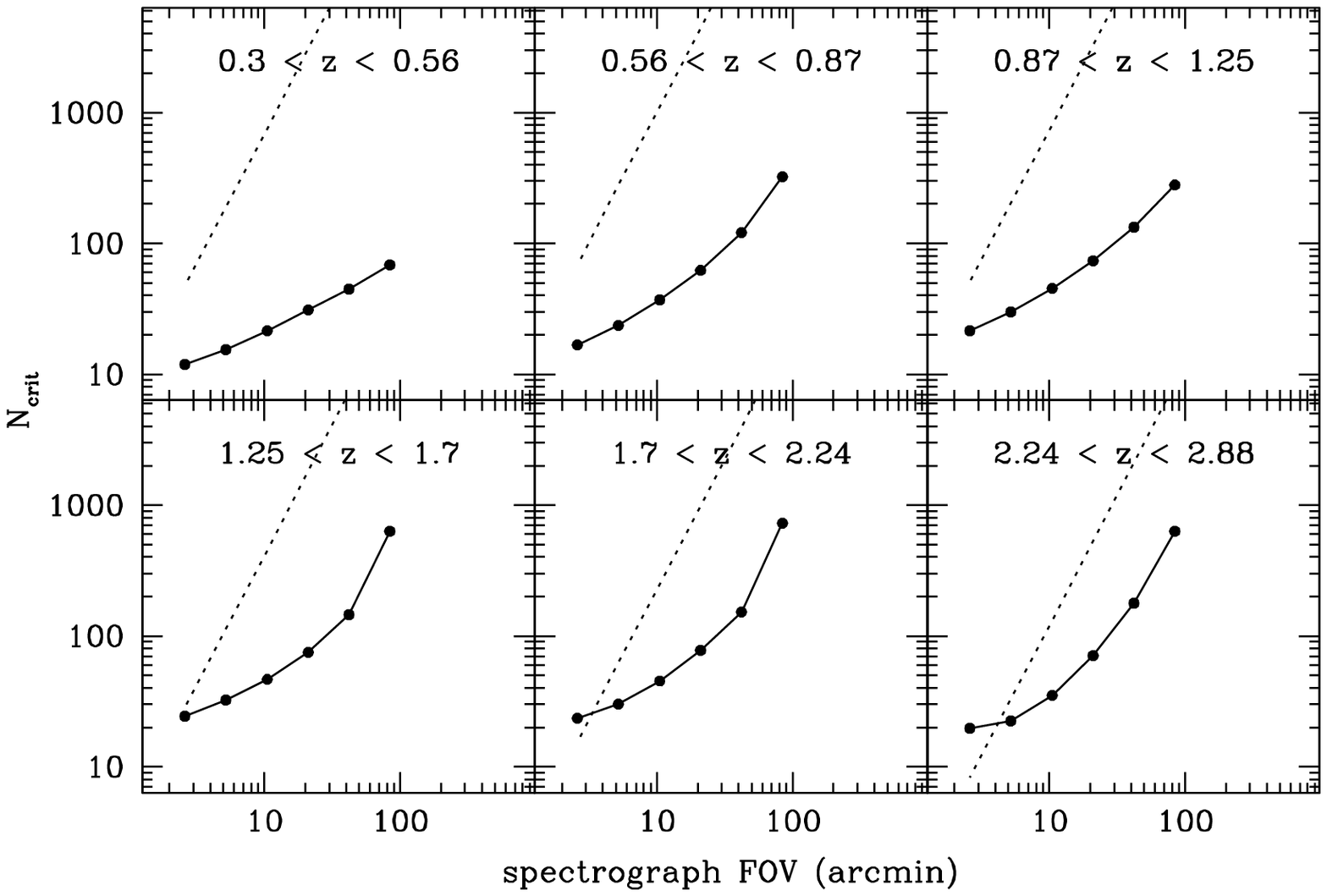}}

\caption{The critical number of redshifts at which the variance in the mean redshift becomes dominated by the effects of large scale structure, or cosmic variance. The solid lines give the derived $ N_{crit}$ for $ I_{AB} < 24.5$ in six representative redshift bins for spectrograph fields of view from a few arc minutes up to 1.4 degree. The dotted line gives the average number of galaxies within the field of view of the spectrograph. The difference between these lines indicates the maximum permitted sampling rate. A fully sampled spectroscopic survey is likely to be severely cosmic variance limited and much lower sampling rates are required to keep the effects of large scale structure comparable to the Poisson term. }
\label{Appendix-1}
\end{figure*}

Figure-\ref{Appendix-1} shows the derived $N_{crit}$ for galaxies with $I_{AB} < 24.5$ in six representative redshift bins for spectrograph fields of view ranging from a few arcmin (e.g. NIRSpec on JWST) up to fields of 1.4 degree, which is approaching the largest that is likely practical on an 8-m class telescope. For reference, the current VIMOS spectrograph on the VLT has a field of about 15 arcmin square. These rather low values of $N_{crit}$ reflect the familiar observational experience that the large scale structure in a given survey field starts to become apparent in the $N(z)$ distribution after only a few galaxies have been observed. 

\begin{figure*}\label{graphics}

{\includegraphics[angle=0,width=12.5cm,height=10cm]{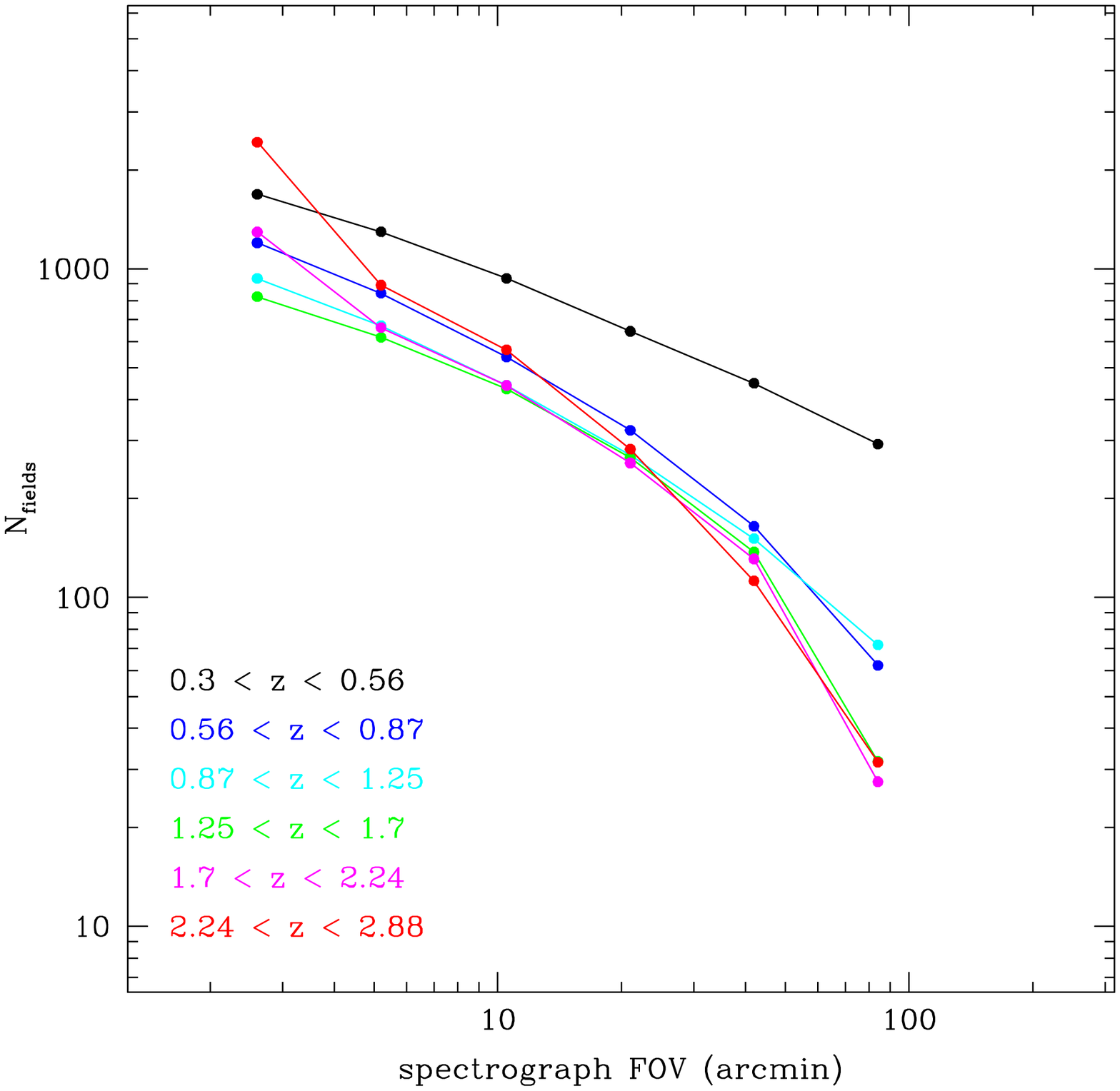}}
\caption{The minimum number of independent fields required in order to ensure that the uncertainty in the mean redshift of a given redshift bin is not dominated by large scale structure or cosmic variance, and is equal to the Poisson variance from $ 10^{4}$ galaxies. Clearly for an all sky survey, the number of required fields will be set by lower redshift range.}
\label{Appendix-2}
\end{figure*}

In particular, it can be seen that the value of $N_{crit}$ is usually very much smaller than the average number of galaxies within the field of view of the spectrograph, which is shown as the dotted line in each panel.  The difference between these curves indicates the maximum permitted sampling rate. This emphasizes that the redshift distribution from a fully sampled spectroscopic survey is likely to be severely cosmic variance limited and that very much lower sampling rates are required to keep the effects of large scale structure comparable to the Poisson term. For example, a survey with the VIMOS spectrograph with a fields of view of about 15 arcminutes square would require a sampling rate at low redshift of 2$\%$ (i.e. one galaxy in 50) or less. Only at the highest redshifts and with the smallest field sizes is cosmic variance unimportant The low sampling rates demanded by this analysis would pose quite severe inefficiencies on the utilization of slit-mask spectrographs for such a program of spectroscopic calibration of photo-z bins.  These would be mitigated for fibre-fed spectrographs, although the performance of these at $I_{AB} \sim 24.5$ has not yet been proven. Regardless, it is clear that the survey fields for such a programme would have to be distributed over a significant portion of the sky.

The point at which $N_{crit}$ starts to increase as the square of the field of view (i.e. where the solid line becomes parallel to the dotted lines) shows the point at which it is safe to mosaic adjacent survey fields to build up survey area and galaxy number. This occurs at about degree-scales for $z > 1.7$.

If we then take the minimum of $N_{crit}$ and the available number of galaxies in the field, which is almost always given by the former, we can then compute the minimum number of independent fields that will be required in order to attain an uncertainty in the mean redshift of this particular redshift bin $\langle z\rangle$ that is equivalent to the Poisson variance from $10^{4}$ galaxies. This is shown in Figure-\ref{Appendix-2}.  

If one imagines doing a single survey to cover the entire redshift range, then the number of fields will generally be set by the lower redshifts, where the effects of large scale structure are most severe. Only at very small field sizes does the low number density of very high redshift galaxies become the limiting factor.  It can be seen that about 400 widely spaced degree-scale survey fields, 700 VIMOS fields, or about 2000 NIRSpec (3-arcminute) survey fields would be required.  This requirement clearly approximates an all-sky sparse sampled survey.

The difficulty of implementing such a scheme was a major motivation for considering the alternative approach to constructing N(z) which is developed in the main paper.  The effects of large scale structure are irrelevant in that approach which considers photo-z calibration on an object-by-object level.

\end{document}